\documentclass[12pt]{article}

\usepackage{newtxtext,newtxmath}
\usepackage[utf8x]{inputenc}
\usepackage[english]{babel}
\usepackage{amsmath,amsbsy,amsfonts}
\usepackage{graphicx}
\usepackage{subfig}
\usepackage{color}
\usepackage{float}
\usepackage{cite}
\usepackage[left=1cm,right=1cm,top=2cm,bottom=2cm]{geometry}
\usepackage{physics}
\usepackage[colorlinks=true, allcolors=red]{hyperref}

\numberwithin{equation}{section}

\begin{document}

\title{Modeling displaced squeezed number states in waveguide arrays}

\author{B.M. Villegas-Martínez$^*$, H.M. Moya-Cessa and F. Soto-Eguibar\\
\small Instituto Nacional de Astrofísica, Óptica y Electrónica, INAOE \\
\small {Calle Luis Enrique Erro 1, Santa María Tonantzintla, Puebla, 72840 Mexico}\\
$^*$\small {Corresponding author: bvillegas@inaoep.mx}}

\date{\today}

\maketitle

\begin{abstract}
We present an exact analytical solution for a one-dimensional zigzag waveguide array with first and second neighbor interactions. It is found that the waveguide system possess a classical analog to the displaced squeezed number states. The exact solution was compared directly with the numerical solution showing a perfect agreement between both results. The implication of a linear index of refraction changing as a function of the site number is also studied. In this case, we show that the first neighbor interaction strongly influences the periodicity of Bloch oscillations.
\end{abstract}

\section{Introduction}
Optics and Photonics basic elements are so well theoretically understood and tested in experiments that provide a quite powerful and simple laboratory tool where it is possible to underline quantum-optical analogies. In principle, this formulation derives from the mathematical isomorphism between the temporal Schrödinger for a free quantum particle and the paraxial wave-equation, describing spatial light propagation in photonic guiding structures, namely, photonic lattices. The successful adoption of this classical system composed of $N$ evanescent coupled optical waveguides, has inspired many researchers to simulate and observe a wide variety of quantum processes which include quantum Zeno effect\cite{1A,1.1A,1.2A}, optical Bloch oscillations\cite{2A,2.1A, 2.2A, 2.3A}, Anderson localization \cite{3A,3.1A,3.2A,3.3A} and Rabi oscillations\cite{4A,4.1A,4.2A} among others\cite{5A,5.1A,5.2A,5.3A,5.4A}. In spite of light propagation along waveguides is nowadays well known, it is usually assumed in most models that the couplings among near waveguides only take place between first nearest neighbors and the second-order coupling is not considered. Nevertheless, in certain circumstances, the influence of a second interaction has remarkable relevance in some systems, such as in biomolecules \cite{6A} and energy transfer in polymer chains \cite{7A}. Albeit this next-nearest-neighbor interaction can be introduced by the use of two layers complex structures in a zigzag waveguide arrangement\cite{8A,9A,10A}. Nonlinear properties of the semi-infinite and symmetric arrangement of the lattice itself, has been used in several theoretical as well experimental work \cite{8A,11A,12A}. In the linear regime case, Mahdavi and coworkers recently demonstrated that an index difference on the individual guides in a zigzag structure is an ideal environment to reach to the squeeze Bloch oscillations in the space domain. Indeed, the effect of neglecting the linear gradient index in the waveguides provides a classical analog to the squeezed number and squeezed coherent intensity distribution in quantum optics\cite{13A}. These theoretical outcomes went beyond with its study in a sinusoidal bent squeezed-like lattice\cite{14A}.
Along the same line, the analysis developed in Ref.\cite{15A} is shown that propagation of classical light in arrays of specially modulated coupled optical waveguides may simulate quantum processes of two-mode squeezing in nonlinear media. Interestingly, the theoretical possibility to explore squeezing in semi-infinite and asymmetric zigzag photonic lattices was based on a well-defined coupling configuration, where it was possible neglect all nearest-neighbor interactions between two adjacent layers and the strength of coupling was determined by the next-nearest neighbors at the same layer. The above, in turn begs an open-ended question: how affect influences of the first and second-order neighbor interactions to the squeezed light Bloch oscillations?. \\
In this contribution, we cover the remaining situation not taken into account in Ref\cite{13A} as well as extending the generalization where the waveguide array admits an exact analytical solution. We have to mention at this point, that the light intensity distribution of this waveguide array with particular interaction constants up to second neighbors, it is equivalent to the photon number distribution corresponding to so-called displaced squeezed number state. To the best of our knowledge, classical analogous of a displaced squeezed number state has not yet been investigated and this is our aim in the present work.

\section{The model}
In order to perform the task, we use a generalization of the model introduced in Ref. \cite{13A}, namely, a derivation based on the results in Refs.  \cite{11A,13.1A,20}. The system consists of asymmetric and semi-infinite zigzag photonic lattice, where the propagation constant is proportional to  the waveguide number in the array, taking into account the interaction between neighbors up to second-order. Therefore, according to the coupled mode theory, the discrete coupled set of equations for the light propagation in each of the waveguides described in this system may be written as 
\begin{equation} \label{1}
i \frac{d}{dz} \mathcal{E}_n(z) + \mu_n \mathcal{E}_n(z) + \alpha_1 \left[C_n^{(1)}  \mathcal{E}_{n-1}(z) + C_{n+1}^{(1)} \mathcal{E}_{n+1}(z) \right] + \alpha_2 \left[C_{n}^{(2)} \mathcal{E}_{n-2}(z) + C_{n+2}^{(2)} \mathcal{E}_{n+2}(z) \right]=0,\quad
n=0,1,2,...
\end{equation}
where $\mathcal{E}_{n}(z)$ is the electric field amplitude in the $n$-th waveguide ($n=0,1,2,...$), being $ \mathcal{E}_{n}(z)=0 $ for $n<0$. The first term in the above expression corresponds to the change of $ \mathcal{E}_{n}(z)$ along the propagation distance $z$. In the second term, $\mu_{n}$ represents the propagation constant of each waveguide; in the above model, we assume that each waveguide has the same propagation constant $\mu$ with a linear transverse gradient, i.e. $\mu_{n}=\mu + \alpha_{0} n$, where $\alpha_{0}$ is a gradient factor which gives rise to Bloch oscillations \cite{2.2A, 16A,18A,19A}. The third and fourth terms represent the first and second neighbor coupling modulated by the functions $\alpha_{1}$ and $\alpha_{2}$, and which for simplicity, we consider as $z$-independent; nevertheless, the $z$-dependence of functions are thoroughly studied in \cite{20}. In the weak coupling regime, the coupling coefficients $C_n^{(1)}$ and $C_n^{(2)}$ depend on the distance between nearest neighbors and next-nearest neighbors, such that $C^{(1)}_{n}=C \exp[-\frac{d^{(1)}_{n}-d_{1}}{\kappa}]$ and $C^{(2)}_{n}=C \exp[-\frac{d^{(2)}_{n}-d_{2}}{\kappa}]$, being $C$ a constant, $\kappa$ a free parameter which is determined from coupled mode theory \cite{21,21.5,22,23}, $d^{(1)}_{n}=d_{1}-\frac{\kappa}{2} \ln(n)$ is the distance between the $n$-th site and its first-order neighbor $(n+1)$-th, and $d^{(2)}_{n}=d_{2}-\frac{\kappa}{2} \ln[n \left(n-1\right)] $ denotes the distance between $n$-th and its second-order neighbors $(n+2)$-th, respectively.\\
We apply the substitution $\mathcal{E}_{n}(z)=\Psi_{n}\left(Z\right) \exp\left(i \mu z\right)$ to Eq.~\eqref{1},  that may be written in the dimensionless form
\begin{align} \label{2}
& i \frac{d}{dZ} \Psi_{n}(Z) + \lambda n  \Psi_{n}(Z) + \alpha_{1}\left[\sqrt{n} \Psi_{n-1}(Z) + \sqrt{n+1} \Psi_{n+1}(Z)\right]  
\nonumber\\ &
+ \alpha_{2}\left[\sqrt{n(n-1)} \Psi_{n-2}(Z) + \sqrt{(n+1)(n+2)} \Psi_{n+2}(Z)\right]=0, \quad n=0,1,2,...
\end{align}
with $\lambda=\alpha_{0}/C$ and $Z=C z$ are normalized quantities.\\
In order to find the analytical solution of Eq.~\eqref{2}, we adopt a method already known from  quantum optics to deal with classical optics problems; the essence of this method consists in defining annihilation and creation operators for the waveguide number basis, i.e. $\hat{a}\ket{n}=\sqrt{n}\ket{n-1}$ and $\hat{a}^{\dagger}\ket{n}=\sqrt{n+1}\ket{n+1}$, where $\ket{n}$ represents the optical analogue of Fock states. In this case, the differential set Eq.~\eqref{2} can be written as the Schrödinger-like equation 
\begin{equation} \label{3}
i\frac{d}{dZ} \ket{\psi\left(Z\right)}=-\left[ \lambda \hat{n} + \alpha_{1} \left(\hat{a}^{\dagger} + \hat{a}\right) + \alpha_{2} \left(\hat{a}^{\dagger2}  + \hat{a}^2\right)  \right] \ket{\psi\left(Z\right)},
\end{equation}
where $\ket{\psi\left(Z\right)}=\sum_{n=0}^{\infty} \Psi_{n}(Z) \ket{n}$, being  $\Psi_{n}(Z)$ the rescaled field amplitude which depends on the dimensionless propagation distance and where the orthogonality of the waveguide eigenmodes $\delta_{m,n}=\braket{m}{n}$ is satisfied. One can easily prove that if the proposal $\ket{\psi\left(Z\right)}$ is substituted into Eq.~\eqref{3}, the system given by Eq.~\eqref{2} is recovered; therefore, we can solve Eq.~\eqref{3} instead of Eq.~\eqref{2}. Moreover, we observe that Eq.~\eqref{3} contains two subalgebras, the generators $\hat{K}^{+}=\frac{\hat{a}^{\dagger2}}{2}$, $ \hat{K}^{0}=\frac{\left(\hat{n} + 1/2\right)}{2}$ and $ \hat{K}^{-}=\frac{\hat{a}^{2}}{2}$,  constitute a su(1,1) Lie algebra and the generators $\lbrace\hat{n} + 1/2, \hat{a}^{\dagger}, \hat{a}, \hat{I} \rbrace$ correspond to the single-photon algebra; these six components generate the two-photon algebra which obey the commutation relations
\begin{align} \label{4}
\left[\hat{K}^{+}, \hat{K}^{-} \right]=&-2\hat{K}^0, & \left[\hat{K}^{0}, \hat{K}^{+} \right]=&\hat{K}^{+}, & \left[\hat{K}^{0}, \hat{K}^{-} \right]=&- \hat{K}^{-}, \nonumber \\
\left[\hat{K}^{+}, \hat{a}^{\dagger} \right]=&0, & \left[\hat{K}^{+}, \hat{a} \right]=&-\hat{a}^{\dagger}, & \left[\hat{K}^{0}, \hat{a}^{\dagger} \right]=&\frac{\hat{a}^{\dagger}}{2}, \nonumber \\
\left[\hat{K}^{-}, \hat{a} \right]=&0, &  \left[\hat{K}^{-}, \hat{a}^{\dagger} \right]=& \hat{a}, & \left[\hat{K}^{0}, \hat{a} \right]=&-\frac{\hat{a}}{2}. 
\end{align}

\section{Solution of the Schrödinger type equation}
\begin{figure}[H]
	\centering
	{\includegraphics[width=0.68\textwidth]{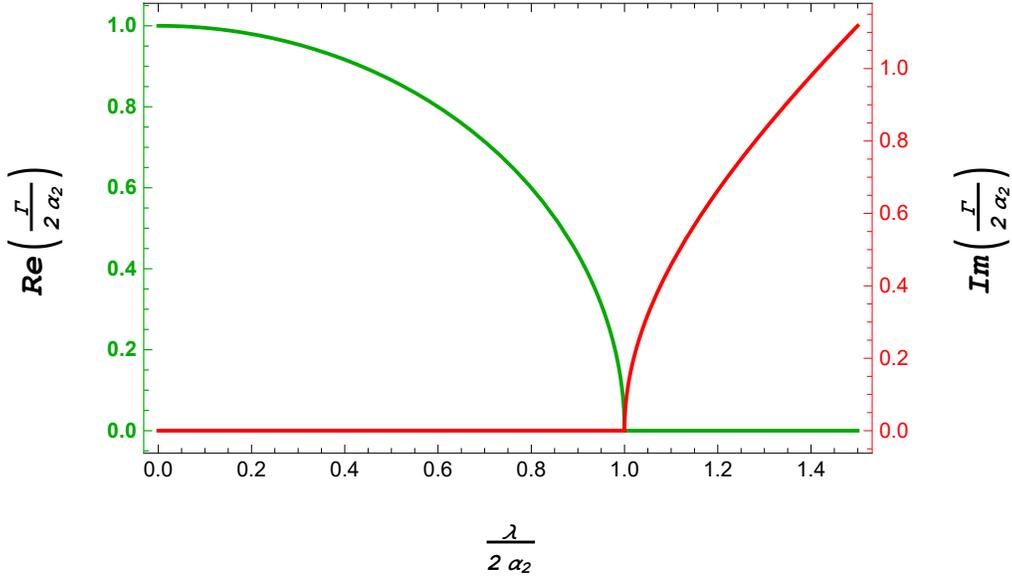}}
	\caption{Real and imaginary values of $\frac{\Gamma}{2\alpha_{2}}$ versus $\frac{\lambda}{2\alpha_{2}}$. Here $\Gamma$ exhibits three different regimes depending on whether $\lambda$ is less, greater of equal to $2\alpha_{2}$.}
	\label{f1}
\end{figure}

\begin{figure}[H]
\subfloat[Numerical solution of Eq.\eqref{2}, carried out with 200 waveguides and solving by the Runge-Kutta-Fehlberg Method.]
{\includegraphics[width=0.42\textwidth]{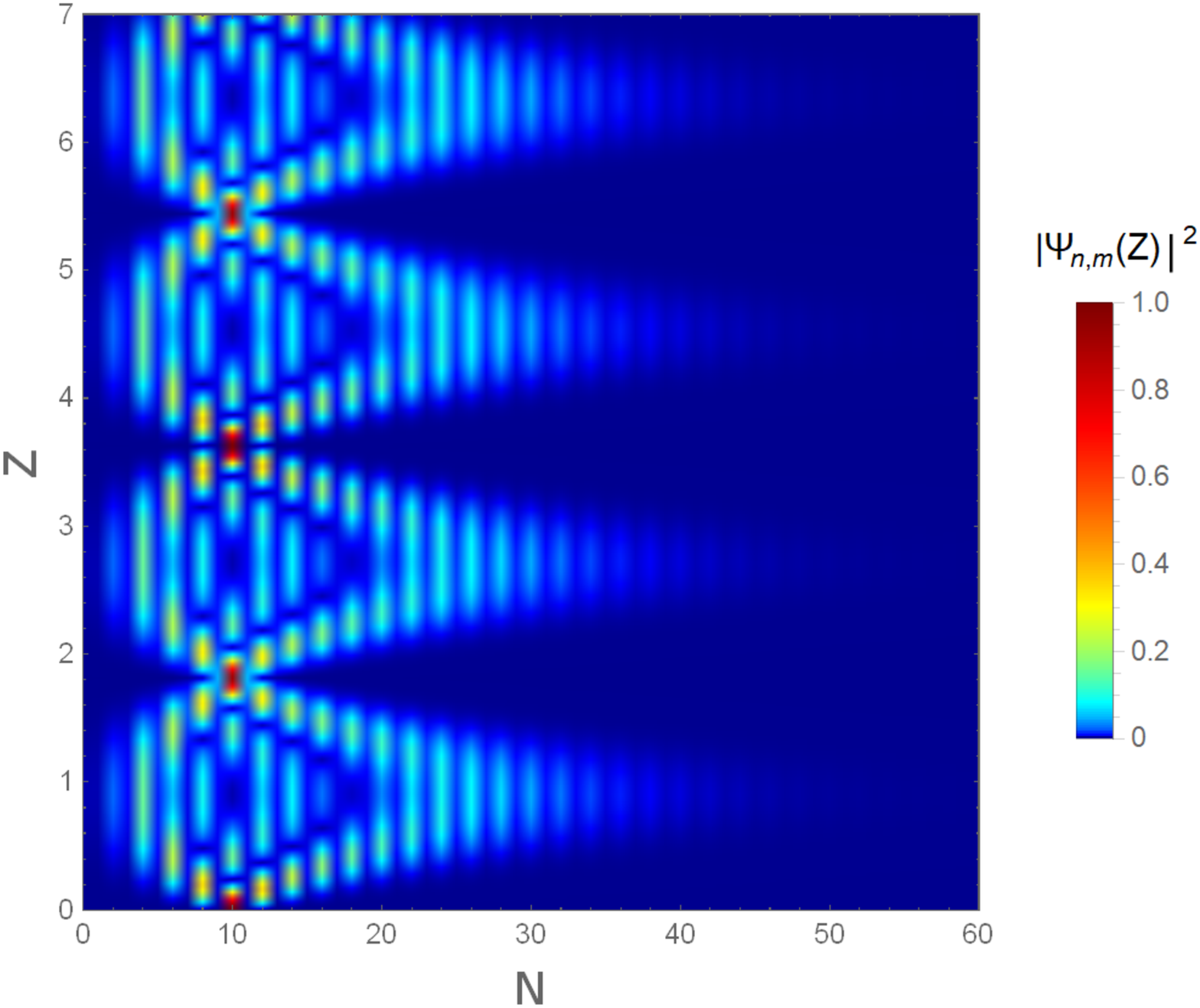}}
\subfloat[Exact analytical solution of Eq.\eqref{20}.]
{\includegraphics[width=0.42\textwidth]{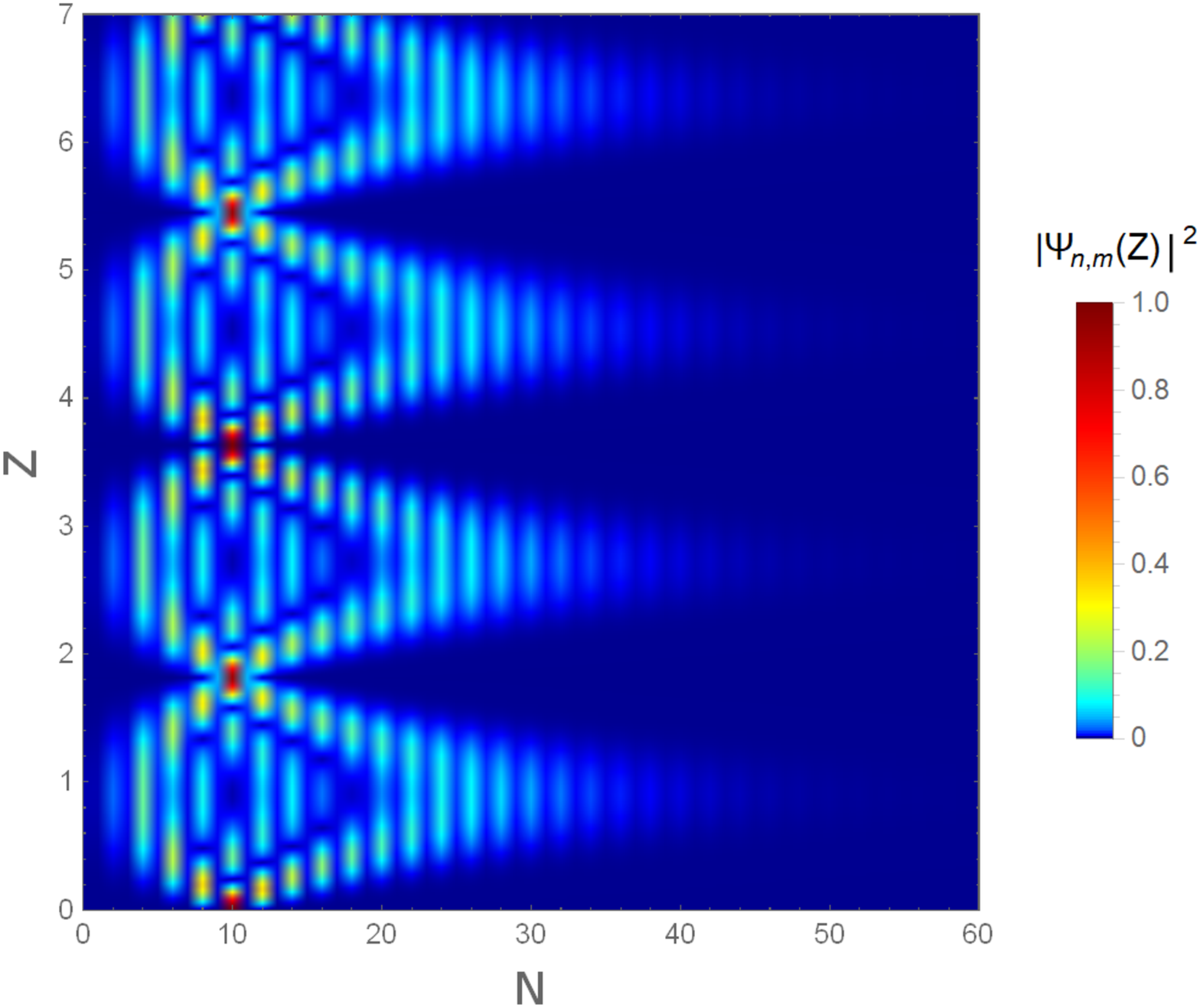}}\\
\subfloat[Comparison between the exact and numerical solution at $n=10$.]
{\includegraphics[width=0.42 \textwidth]{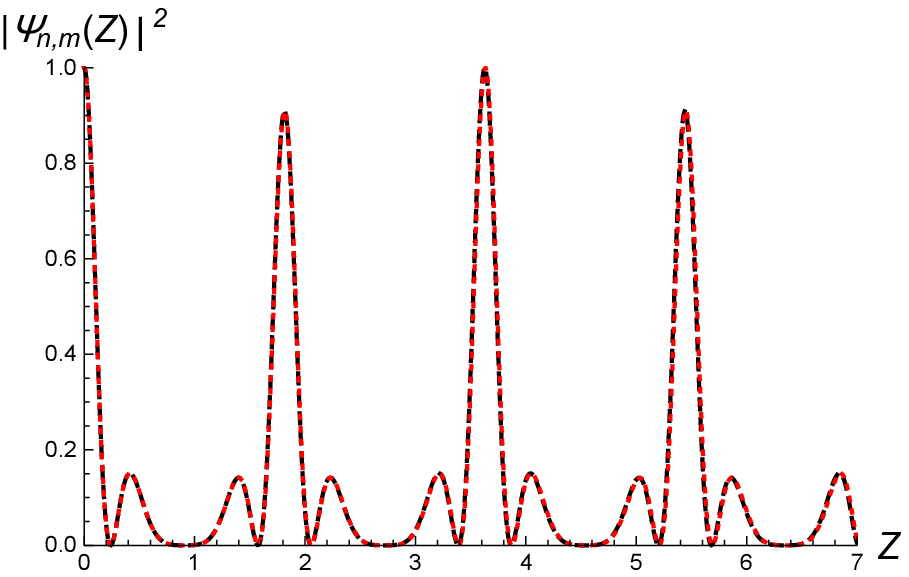}}
\subfloat[Comparison between the exact and numerical solution at $Z=2.6$.] 
{\includegraphics[width=0.42 \textwidth]{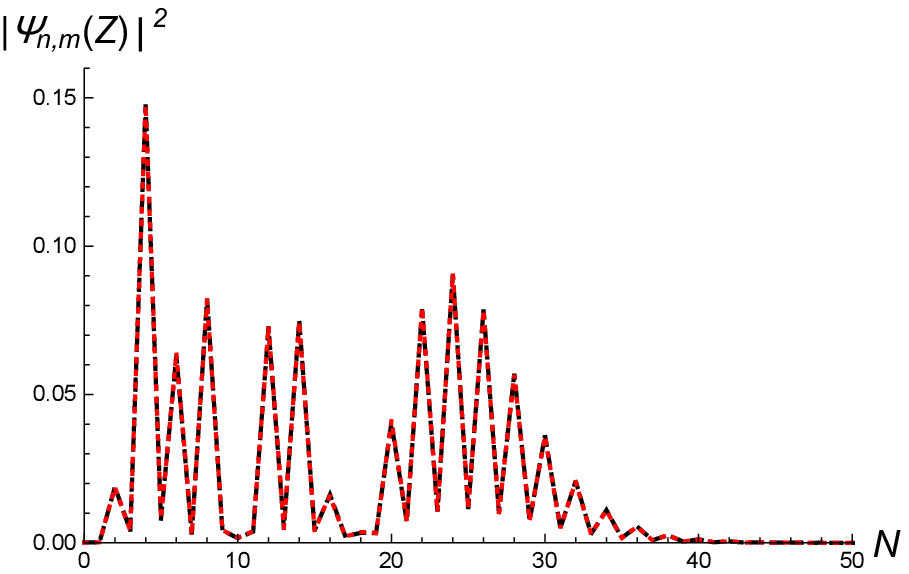}}
\caption{
	Evolution of light intensity $\abs{\Psi_{n,m}(Z)}^2$ for the exact (a) and numerical (b) solution with an initial excitation at $n=10$. In Figs.(c) and (d) our exact solution, denoted by red circles, is validated by comparison with the numerical solution, indicated by the black dashed line at $n=10$ and $Z=2.6$. This comparison shows an excellent agreement between both results.These plots were generated using the parameters $\lambda=2, \alpha_1=0.1, \alpha_2=0.5$.}
\label{f2}
\end{figure}

\begin{figure}[H]
	\subfloat[$\alpha_{1}=0.5$.]
	{\includegraphics[width=0.42\textwidth]{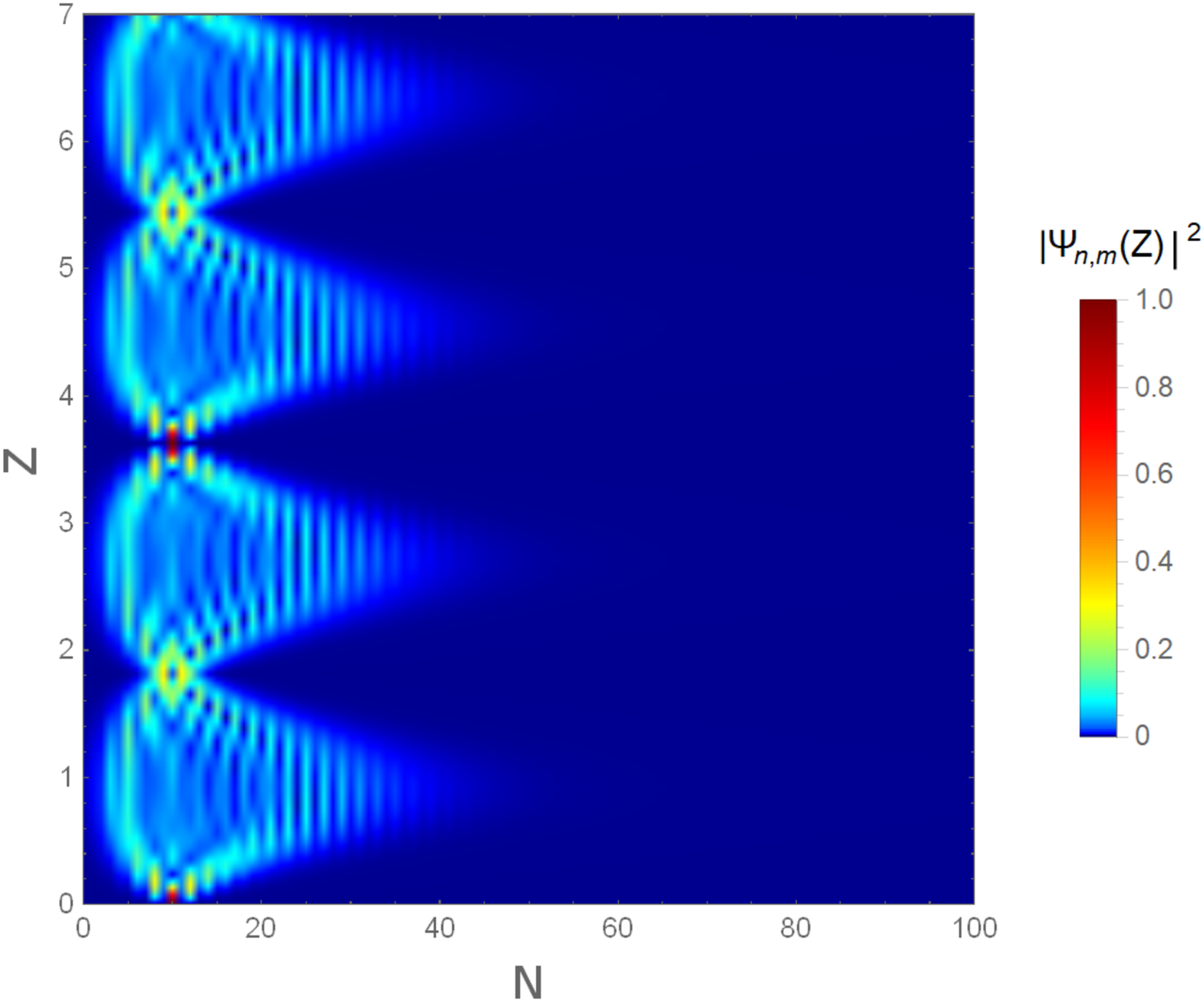}}
	\subfloat[$\alpha_{1}=2$.]
	{\includegraphics[width=0.42\textwidth]{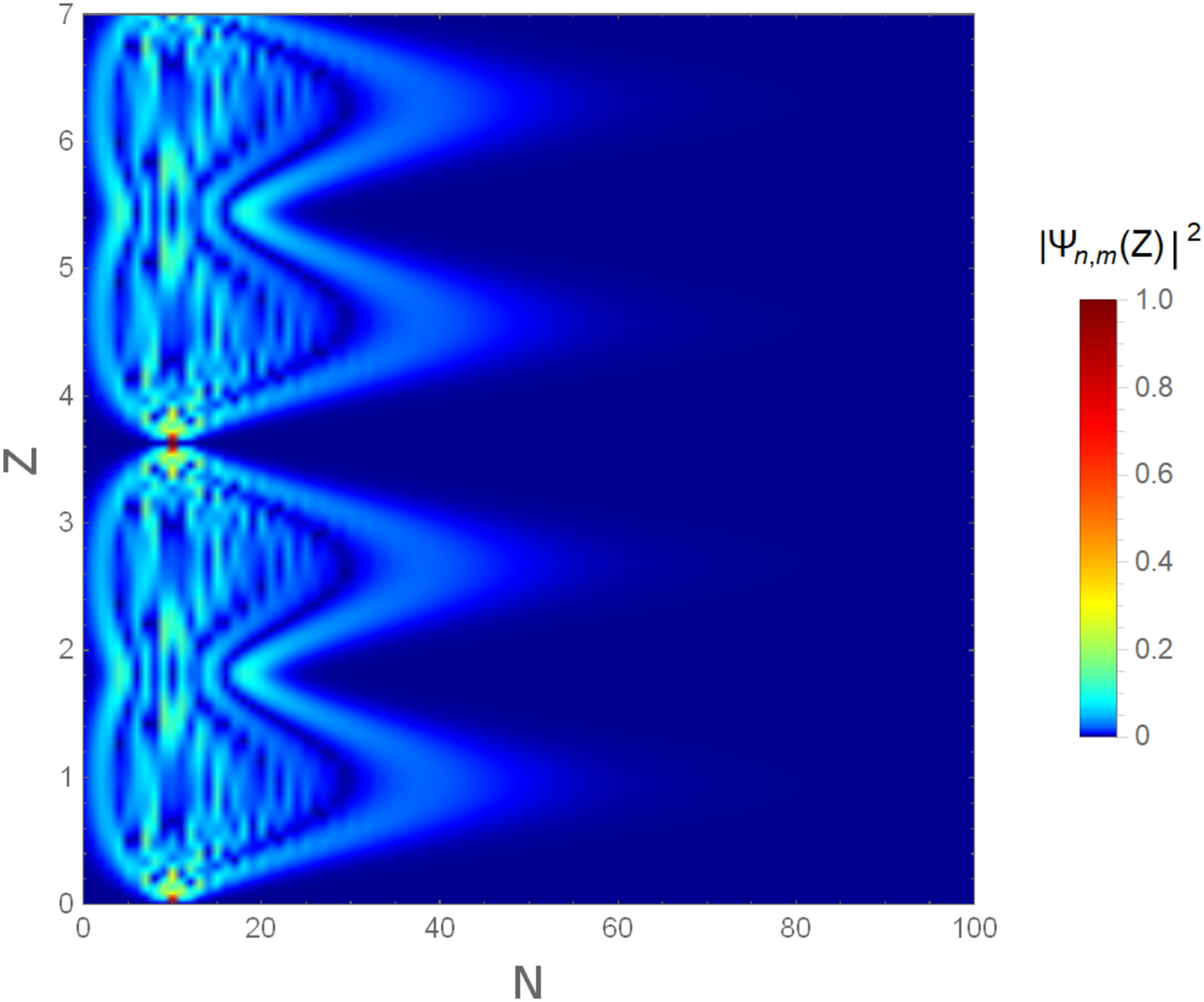}} \\
	\subfloat[$\alpha_{1}=4$.]
	{\includegraphics[width=0.42\textwidth]{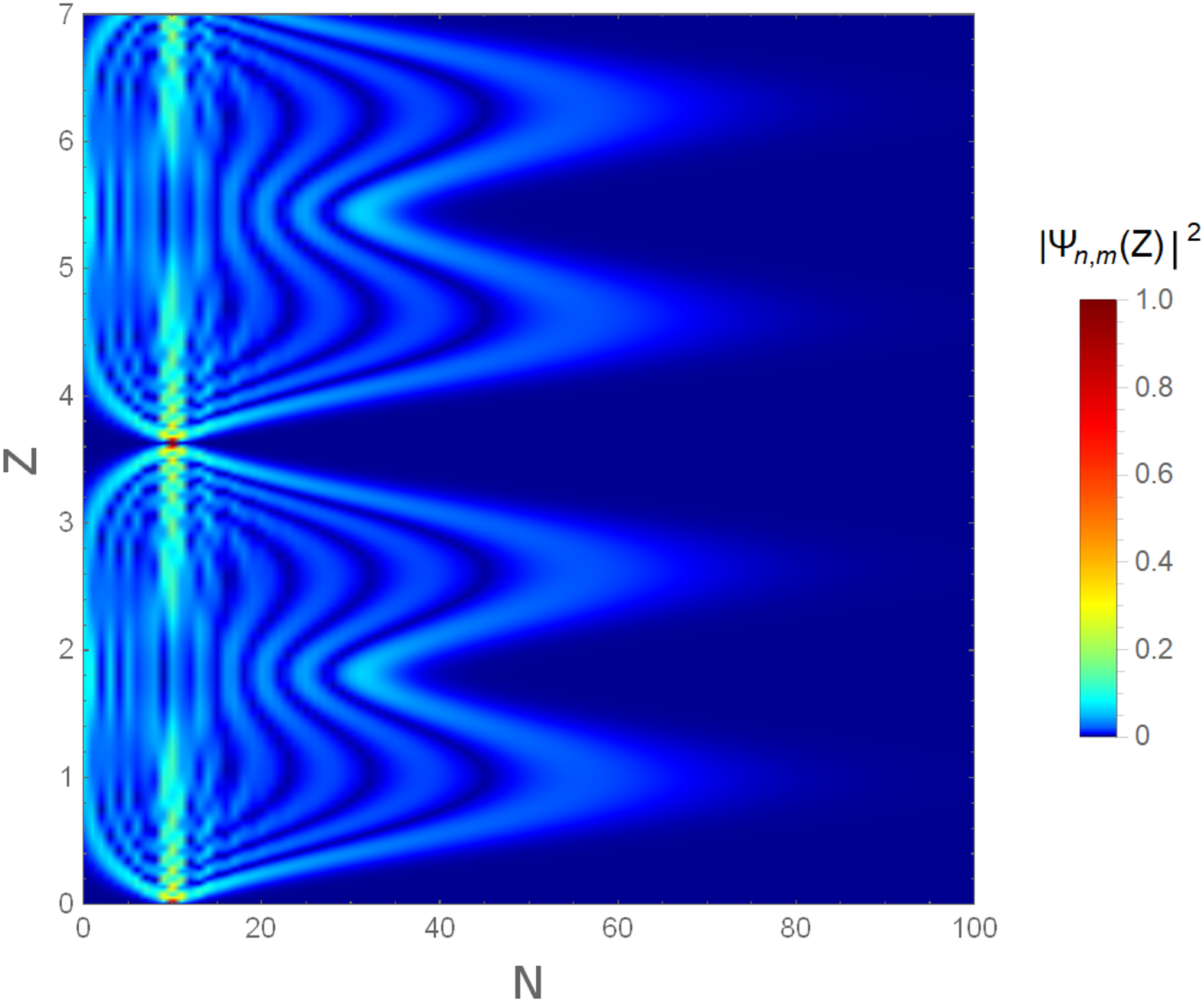}}
	\subfloat[$\alpha_{1}=8$.]
	{\includegraphics[width=0.42\textwidth]{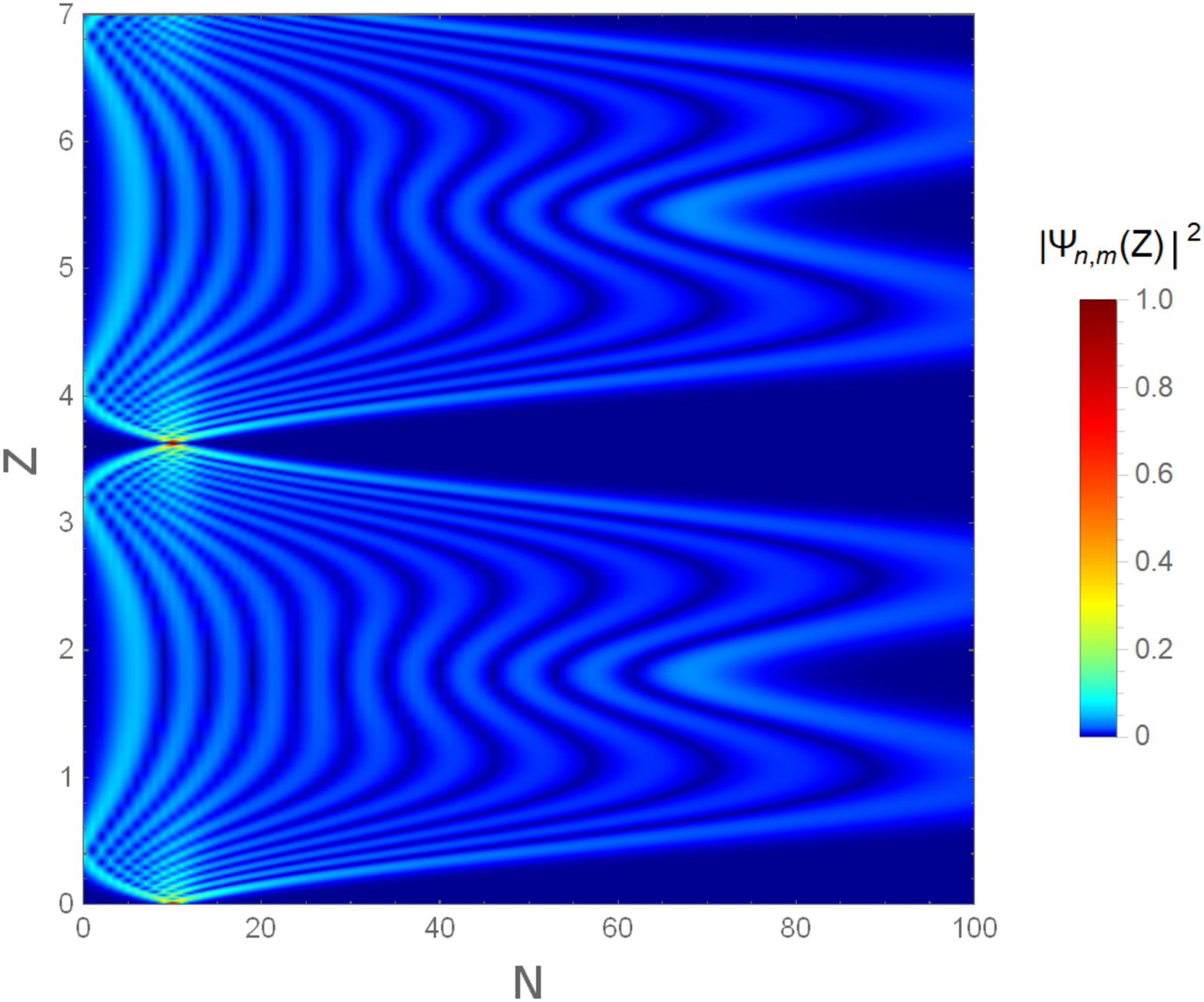}}
	\caption{Light intensity evolution along $100$ waveguides corresponding to an initial excitation at $n=10$, as obtained by the exact solution. In this case the parameters $\lambda=2$ and $\alpha_{2}=0.5$ are fixed whereas the coupling coefficient $\alpha_{1}$ is varied at $\alpha_{1}=0.5,2,4$ and $8$. As $\alpha_{1}$ is increased, the Bloch oscillation period is modified being now twice $Z_{p}$.}
	\label{f3}
\end{figure}

It is noteworthy to mention that, at the outset, we distinguish two different cases to solve the Schrödinger type equation \eqref{3}. The first one is when $\lambda \neq -2\alpha_2$ and second when $\lambda = -2\alpha_2$, the later case leads to the Schrödinger equation of a shifted linear potential that can be solved by using the extended Baker-Campbell-Hausdorff formula \cite{extbchf}. For more details, please see online supplementary file for this article. In the following, we therefore focus on the case $\lambda \neq -2\alpha_2$.\\
Let us consider the transformation $\ket{\Phi(Z)}=\hat{D}\left(\beta\right) \ket{\psi\left(Z\right)}$ in Eq~\eqref{3}, where $\hat{D}(\beta)=\exp\left(\beta \hat{a}^\dagger-\beta^{*}\hat{a}\right)$ is the Glauber displacement operator\cite{24,25}. If we restrict the displacement parameter $\beta$ to real numbers and use the commutation relations \eqref{4}, Eq.~\eqref{3} becomes
\begin{align} \label{5}
i \frac{d}{dZ} \ket{\Phi(Z)}=& \Bigg\lbrace-2 \alpha_{2} \left( \hat{K}^{+} + \hat{K}^{-} + \frac{\lambda}{\alpha_{2}} \hat{K}^{0} \right) + \Big[\beta\left(\lambda + 2 \alpha_{2}\right)-\alpha_{1}\Big] \left(\hat{a} + \hat{a}^{\dagger}-\beta\right)  + \alpha_{1}\beta + \frac{\lambda}{2}\Bigg\rbrace \ket{\Phi(Z)};
\end{align}
with the choice $\beta=\frac{\alpha_{1}}{\lambda + 2 \alpha_{2}}$, under the restriction $\lambda+2\alpha_2\neq0$, the term $\hat{a} + \hat{a}^{\dagger}-\beta$ is canceled, which allows us to write the formal solution to this equation as
\begin{align} \label{6}
\ket{\Phi\left(Z\right)}=&\exp\left[-i \frac{2 \alpha^2_{1} + \lambda \left(\lambda  + 2 \alpha_{2} \right)}{2 \left(\lambda + 2 \alpha_2 \right)} Z\right]  \exp\left( 2 i \alpha_{2} \hat{H} Z \right) \ket{\Phi(0)},
\end{align}
where
\begin{equation}\label{7}
\hat{H}= \hat{K}^{+} +  \frac{\lambda}{\alpha_{2}} \hat{K}^{0} + \hat{K}^{-}  .
\end{equation}
The solution $\ket{\psi(Z)}$ of Eq.~\eqref{3} is recovered using the inverse transformation $\ket{\psi\left(Z\right)}=\hat{D}^{\dagger}\left(\beta\right) \ket{\Phi\left(Z\right)}$, to give
\begin{equation} \label{8}
\ket{\psi\left(Z\right)}= \exp\left[-i \frac{2 \alpha^2_{1} + \lambda \left(\lambda  + 2 \alpha_{2} \right)}{2 \left(\lambda + 2 \alpha_2 \right)} Z\right] \hat{D}^{\dagger}\left(\beta\right)  \exp\left( 2 i \alpha_{2} \hat{H} Z \right) \hat{D}\left(\beta\right) \ket{\psi(0)}.
\end{equation}
We write the identity operator as $\hat{I}= \exp\left(2 i \alpha_{2} \hat{H} Z \right)  \exp\left(-2 i \alpha_{2} \hat{H} Z \right)$, insert it in the above expression and use that
\begin{align} \label{9}
\exp\left(-2 i \alpha_{2} \hat{H} Z \right)  \hat{D}^\dagger \left(\beta\right)
\exp\left(2 i \alpha_{2} \hat{H} Z \right)=\hat{D}^\dagger\left(\beta+\eta\right),
\end{align}
where 
\begin{equation} \label{10}
\eta= \frac{\alpha_{1}}{\left(\lambda + 2 \alpha_{2}\right) \Gamma}\left[2 \Gamma \sinh^2\left( \Gamma Z/2\right) - i \left(\lambda + 2 \alpha_{2} \right) \sinh\left( \Gamma Z\right) \right],
\end{equation}
with $\Gamma = \sqrt{4 \alpha^2_{2}-\lambda^2}$, to write
\begin{equation} \label{11}
\ket{\psi\left(Z\right)}= \exp\left[-i \frac{2 \alpha^2_{1} + \lambda \left(\lambda  + 2 \alpha_{2} \right)}{2 \left(\lambda + 2 \alpha_2 \right)} Z\right] 
\exp\left( 2 i \alpha_{2} \hat{H} Z \right)
\hat{D}^\dagger \left(\beta+\eta\right)
\hat{D}\left(\beta\right)  
\ket{\psi(0)}.
\end{equation}
To further simplify the above expression for $\ket{\psi\left(Z\right)}$, we use the displacement operator identity $\hat{D}\left(\chi\right) \hat{D}\left(\gamma\right)=\exp\left(\frac{\chi \gamma^{*}-\chi^{*}\gamma}{2}\right) \hat{D}\left(\chi+\gamma\right)$ and we obtain
\begin{equation} \label{12}
\ket{\psi\left(Z\right)}=\exp\left(-\frac{\nu}{2}\right)
\exp\left( 2 i \alpha_{2} \hat{H} Z \right)
\hat{D}^\dagger\left(\eta\right) 
\ket{\psi\left(0\right)},
\end{equation}
being
\begin{equation} \label{13}
\nu=\frac{i}{ \left(\lambda + 2 \alpha_{2} \right)\Gamma} \Bigg\lbrace \left[2 \alpha^2_{1} + \lambda \left(\lambda + 2 \alpha_{2} \right)\right] \Gamma Z- 2 \alpha^2_{1} \sinh\left(\Gamma Z\right) \Bigg\rbrace.
\end{equation}
It may be noticed that with the aid of the two subalgebra structures in \eqref{4}, the exact solution of Eq.~\eqref{3} can be recast as the product of six exponential as follows
\begin{equation} \label{14}
\ket{\psi\left(Z\right)}=
\exp\left(-\frac{\nu}{2}-\frac{\abs{\eta}^2}{2}\right)
\exp\left[g_{1}(Z) \hat{K}^{+}\right] \exp\left[g_{0}(Z) \hat{K}^{0}\right] \exp\left[g_{1}(Z) \hat{K}^{-}\right]
\exp\left(-\eta \hat{a}^{\dagger} \right)  \exp\left(\eta^* \hat{a}\right)\ket{\psi\left(0\right)},
\end{equation}
where
\begin{subequations} \label{15}
\begin{align} 
g_{1}(Z)=&\frac{2i \alpha_{2}\sinh\left(\Gamma Z\right)}{\Gamma \cosh\left(\Gamma Z\right) - i \lambda  \sinh\left(\Gamma Z\right)},    \\
g_{0}(Z)=&-2 \ln \left[\cosh(\Gamma Z)-i \frac{ \lambda}{\Gamma} \sinh(\Gamma Z)\right].
\end{align}
\end{subequations}
It is worth mentioning that with the choice of the parameters $\lambda=0$, $\alpha_{1}=1/\sqrt{2}$ and $\alpha_{2}=1/2$, the above expression turns out to have a formal similarity to the solution of a linear anharmonic repulsive oscillator, where $Z$ plays the role of time \cite{28}, and if we choose $\alpha_{1}=0$, then the system is reduced to the repulsive system alone \cite{28}. Therefore dynamics of both quantum systems can be simulated by the waveguide lattice given in Eq.\eqref{1}.\\ 
An intriguing feature of Eq.~\eqref{14} is that it contains the parameter $\Gamma$ which takes real values when $ \frac{\lambda}{2\alpha_{2}} < 1$ and purely imaginary values when $ \frac{\lambda}{2\alpha_{2}}>1$ (see Fig.\ref{f1}); when $\Gamma$ goes from real to purely imaginary, the hyperbolic functions of our solution changes into trigonometric ones; both behaviors are limited by the phase transition at the critical point $\lambda=2\alpha_{2}$.
These three regimes come from the competition between the discrete light propagation over a large waveguide number, and the Bragg diffraction, where the light comes back to arrangement with smaller propagation constant. Then, the existence of the critical point $\lambda=2\alpha_{2}$ reflects the competition between these two processes in the array, where for a larger waveguide number $n>>1$, both the strength of the propagation constant and coupling coefficient $\alpha_2$ are proportional to $n$. Herein a linear potential $\lambda$ much bigger than the second-order coupling coefficient modulated by $\alpha_{2}$, leads to a discrete diffraction dominance, whereas for the opposite case, when $\lambda>\alpha_{2}$, the Bragg diffraction prevails, ensuring the appearance of squeezed Bloch-like oscillations\cite{13A,14A}. Importantly our general analysis also investigated this kind of Bloch oscillations, but now when both first and second nearest-neighbor interactions are included in the photonic lattices\cite{13A}. In this case, one might expect sudden changes in the Bloch period oscillation given by the the coupling strength $\alpha_{1}$ proportional to $\sqrt{n}$ for large n sites.
In order to observer this phenomena, let us assume that light is injected at $n$-th waveguide, i.e. $\ket{\psi(0)}=\ket{n}$, then the light amplitude in the site $m$-th, $\Psi_{n,m}\left(Z \right)=\braket{m}{\psi(Z)}$, is given by
\begin{align} \label{16}
\Psi_{n,m}\left(Z \right)=& \exp\left(-\frac{\nu}{2}\right) \sum_{k=0}^{\infty} \bra{m} \exp\left( 2 i \alpha_{2}\hat{H} Z \right) \ket{k}\bra{k} \hat{D}^{\dagger}\left( \eta \right) \ket{n}
\end{align}
where we have applied the completeness relation of the waveguide number basis, $\hat{I}=\sum_{k=0}^{\infty} \ket{k}\bra{k}$. The displacement operator matrix elements are well known \cite{vogel},
\begin{equation} \label{17}
d_{m,n}\left(Z\right) \doteq\bra{m}\hat{D}\left(\eta \right)\ket{n}=\exp\left( -\frac{\abs{\eta}^2}{2}\right)
\begin{cases}
\sqrt{\frac{n!}{m!}}\eta^{m-n}L_n^{\left(m-n \right) }\left( \abs{\eta}^2\right), & m\geq n
\\
\sqrt{\frac{m!}{n!}}\left(-\eta^*\right)^{n-m} L_m^{\left(n-m \right) }\left( \abs{\eta}^2\right), & m<n
\end{cases}
\end{equation}
and for $\exp\left( 2 i \alpha_{2}\hat{H} Z \right)$ we have (see Appendix B)
\begin{align} \label{18}
S_{m,k}\left(Z\right)\doteq&\bra{m} \exp\left( 2 i \alpha_{2}\hat{H} Z \right) \ket{k}
\nonumber \\ 
=&\sqrt{m!k!}\left(\frac{g_1}{2} \right) ^{\frac{m+k}{2}} \exp\left(\frac{g_0}{4} \right) 
\sum_{j=0}^{\infty} \Theta\left(m-j \right)  \Theta\left(k-j \right) 
\cos^2\left[\left(m-j \right)\frac{\pi}{2}  \right] \cos^2\left[\left(k-j \right)\frac{\pi}{2}  \right]
\nonumber \\ & \times
\frac{1}{\left(\frac{m-j}{2} \right) ! \left(\frac{k-j}{2} \right) ! j!}\left[ \frac{2}{g_1} \exp\left(\frac{g_0}{2} \right) \right]^j 
\end{align}
with
\begin{equation} \label{19}
\Theta(x)=
\begin{cases}
0, & x<0,\\
1, & x \geq 0.
\end{cases}
\end{equation}
Substitution of Eq.\eqref{18} and \eqref{17} into Eq.\eqref{16} leads to the following closed expression for the amplitude of the field
\begin{equation}\label{20}
\Psi_{n,m}\left(Z \right)=\exp\left( -\frac{\nu}{2}\right)
\sum_{k=0}^{\infty}S_{m,k}\left(Z\right) d_{k,n}\left(Z\right),
\end{equation}
this exact analytic solution is only applicable to the regimes when $\lambda>\alpha_{2}$ and $\lambda<\alpha_{2}$. For the case when
the second-order coupling coefficient and propagation constant are proportional to waveguide site $N$ (for $n\gg1$), when $\lambda=2\alpha_{2}$. In this situation $\Gamma$ is equal to zero and we must  evaluate the limit $\Gamma\rightarrow 0$ in above expression to get
\begin{align} \label{21}
\Psi_{n,m}\left(Z \right)_{|{\Gamma=0}}= \frac{e^{-i\alpha_{2} Z} e^{i\pi/4}}{\sqrt{2\alpha_{2}Z+i}}
\sum_{k=0}^{\infty} \sqrt{k! m!} \left(-\frac{\alpha_{2} Z}{2\alpha_{2} Z + i}\right)^{\frac{k+m}{2}} d_{k,n}\left(i\alpha_{1} Z\right) 
\nonumber\\ \times
\sum_{j=0}^{\infty} \left(-\frac{e^{i\frac{\pi}{2}}}{\alpha_{2} Z}\right)^{j} \frac{\Theta\left(m-j \right)  \Theta\left(k-j \right) 
\cos^2\left[\left(m-j \right)\frac{\pi}{2}  \right] \cos^2\left[\left(k-j \right)\frac{\pi}{2}  \right]}{j! \left(\frac{k-j}{2}\right)! \left(\frac{m-j}{2}\right)!}.
\end{align}
Figs.\eqref{f2} depict the comparison of light intensity propagation, $I(Z)=\abs{\Psi_{n,m}\left(Z \right)}^2$, obtained by numerically solving Eq.\eqref{2} and by evaluating the analytical exact solution \eqref{20}, in the regime $\lambda>\alpha_{2}$. Figs \eqref{f2} (a) and \eqref{f2} (b) reveal how the input light at $n=10$ spreads and diffracts throughout the array, after then it refocuses at $Z \approx 1.81$ where the optical Bloch oscillation emerges up again. The oscillation period is directly determined by $Z_{p}=\frac{\pi}{\sqrt{\lambda^2-4\alpha^2_{2}}}$, and consistent with Ref\cite{13A}. The intensity profile versus dimensionless propagation distance $Z$ at $n=10$ and the intensity profile versus waveguide number at $Z=2.6$ are shown in Figs \eqref{f2} (c) and  \eqref{f2} (d), where the analytical solution (red circles) reproduces with high accuracy the numerical solution denoted by the black dashed line, here  Eq.\eqref{2} has been solved using the Runge-Kutta-Fehlberg Method. It is important to remark that, although these results came from the choice of parameters $\lambda=2$, $\alpha=0.1$ and $\alpha_{2}=0.5$. The exact solution has been successfully evaluated to a diverse range of values versus the numerical results with excellent agreement in all comparisons, under the restriction $\lambda \geq 0$, $\alpha_{1} \neq 0$ and $\alpha_{2}  \neq 0$. In view of this remark, we present in section 3 the cases when $\lambda \geq 0$, $\alpha_{1} =0$ and $\alpha_{2} \neq 0$ and when $\lambda \geq 0$, $\alpha_{1} \neq 0$ and $\alpha_{2}=0$.
As we already discussed, we are interested on the effect of the first neighbor interaction in the optical squeeze Bloch oscillation. In principle, we expect that this interaction modifies the oscillation period for any input state. In order to corroborate this hypothesis, we plotted different values of $\alpha_{1}$ keeping the same fixed values of $\lambda$ and  $\alpha_{2}$ from Fig.\ref{f2}. We can appreciate in Figs \ref{f3} (a), \ref{f3} (b) and \ref{f3} (c), that the light expands quickly over more numbers of guides as long as $\alpha_{1}$ is increased, herein, the Bloch oscillations begin to couple in pairs and after a certain value of $\alpha_{1}$, such combination produces a new Bloch oscillation pattern with a spatial period of oscillation $2Z_{p}$, as illustrated in Fig \ref{f3} (d). 

\section{Displaced squeezed number states}
In what follows, we show how the mathematical model of Eq.\eqref{2}, with $\lambda\geq 0$, provides a fertile ground for three kinds of scenarios depending on the convenient consideration of first order interaction parameter, second one or both together. The first case, $\lambda>0$, $\alpha_{1} \neq 0$ and $\alpha_{2} =0$  in Eq.\eqref{3}, leads to the following solution
\begin{equation} \label{22}
\Psi_{n,m}\left(Z \right)=\exp \Big\lbrace -i\frac{\alpha^2_{1}}{\lambda^2} \left[\lambda Z-\sin(\lambda Z)\right] + i \lambda m Z \Big\rbrace  d_{m,n} [\eta_{|_{\alpha_{2}=0}}(Z)].
\end{equation}
From this expression is also possible to have Bloch-like revivals with spatial period $Z_{p}=\frac{2\pi}{\lambda}$, such observation of these revivals were theoretically and experimentally reported in\cite{21.5}. Indeed, one can show that if $\lambda=0$ and $\alpha_{1}=1$, we obtain the simple solution $\Psi_{n,m}\left(Z \right)= d_{m,n} (i Z)$. In this model, the light intensity distribution is  equivalent to the photon number distribution of the displaced number states \cite{21}. Next, we turn to the case when $\lambda>0$, $\alpha_{1}=0$ and $\alpha_{2}  \neq 0$ in Eq.\eqref{3} which yields to
\begin{equation} \label{23}
\Psi_{n,m}\left(Z \right)=\exp\left(-i\frac{\lambda}{2} Z \right) S_{m,n} [\eta_{|_{\alpha_{1}=0}}(Z)],
\end{equation}
the above solution gives rise to squeezed light Bloch oscillations and if $\lambda=0$ and $\alpha_{2}=1$, the solution of above model turns out to be $\Psi_{n,m}\left(Z \right)= S_{m,n}(-2i Z)$, where we have a squeeze amplitude twice of the dimensionless propagation $Z$ with phase of $-\pi/2$. Here, the light intensity spatial distribution is equivalent  the photon number distribution for squeeze number states\cite{13A}. Moreover, we can model classically the squeezed vacuum photon distribution when light is injected at site $\ket{n}=\ket{0}$.\\
Scenario 3 for $\lambda>0$, $\alpha_{1} \neq0$ and $\alpha_{2} \neq0$ has already been presented in the previous section. However, one benefits from the exact analytical solution, Eq.\eqref{12},  that under the assumption of $\lambda=0$, the solution, $\ket{\psi(Z)}=\exp\left(-\frac{\eta}{2}\right) \hat{S}\left(-2i\alpha_{2} Z\right)  \hat{D}\left(\eta\right) \ket{\psi(0)}$ may be obtained, which is a product of squeezed and displacement operator that after application to the initial condition $\ket{\psi(0)}=\ket{n}$, we may find that its structure is equivalent to the displaced squeezed number states $\ket{\psi(Z)}=\ket{n,\eta,-2 i \alpha_{2} Z}$. In this case, the light intensity distribution can be calculated by the expression
\begin{equation}
\abs{\Psi_{n,m}(Z)}^2 =\abs{\left\langle m| n,\eta,-2 i \alpha_{2} Z\right\rangle}^2=\abs{\sum_{k=0}^{m} \sqrt{\frac{n! k! \left[\frac{i}{2}\tanh\left(2\alpha_{2} Z \right)\right]^{m+k}}{\cosh\left(2\alpha_{2} Z\right)}} e^{-\frac{\abs{\eta}^2}{2}} \eta^{m-n} L_n^{\left(m-n \right) }\left( \abs{\eta}^2\right) F(2\alpha_{2} Z,k,m)}^2 ,
\end{equation}
for $m\geq n$, while for $m<n$ we have
\begin{equation}
\abs{\Psi_{n,m}(Z)}^2 =\abs{\left\langle m| n,\eta,-2 i \alpha_{2} Z\right\rangle}^2=\abs{\sum_{k=0}^{m} m! \sqrt{\frac{k! \left[\frac{i}{2} \tanh\left(2\alpha_{2} Z\right)\right]^{m+k}}{n! \cosh\left(2\alpha_{2} Z\right)}} e^{-\frac{\abs{\eta}^2}{2}} \left(-\eta^{*}\right)^{n-m} L_m^{\left(n-m \right) }\left( \abs{\eta}^2\right) F(2\alpha_{2} Z,k,m)}^2 
\end{equation}
with
\begin{subequations}\
\begin{align}
F(2\alpha_{2} Z,k,m)&=\sum_{j=0}^{\infty} \Theta\left(m-j \right)  \Theta\left(k-j \right) 
\cos^2\left[\left(m-j \right)\frac{\pi}{2}  \right] \cos^2\left[\left(k-j \right)\frac{\pi}{2}  \right] 
\frac{\left[ -\frac{2 i}{\sinh\left(2\alpha_{2} Z \right) } \right]^j }{\left(\frac{m-j}{2} \right) ! \left(\frac{k-j}{2} \right) ! j!}
\\ 
\eta&=\frac{\alpha_{1}}{2\alpha_{2}}\left[2 \sinh^2\left(\alpha_{2} Z\right)-i \sinh\left(2\alpha_{2} Z\right)\right],
\end{align}
\end{subequations}
and which is equivalent to the displaced squeezed photon number distribution\cite{29,30,31}. Therefore, the classical analogies of displaced Fock states, squeeze number states and squeeze vacuum state could be seen as three limiting cases of classic analogy of the displaced squeezed number states.

\section{Conclusions}
We have obtained a general exact analytical solution for the one dimensional zigzag discrete waveguide array with first and second-neighbor coupling interaction. In all cases, our exact solution shows a good agreement with the numerical results. Indeed, it is clearly lit up for us that the inclusion of a linear index of refraction changing as a function of the site gives rise to Bloch oscillations, always that condition $\lambda>2\alpha_{2}$ is satisfied. Notably, the coupling between diagonal nearest neighboring waveguides, $\alpha_{1}$, plays an important role  to doubling the Bloch oscillation period when only a single site is excited. Such effect shows a crucial difference to the nearest-neighbor model reported in \cite{13A}. On the other hand, our analytical solution on zigzag model can be used to disclose some particular quantum analogies by appropriately tuning the parameters $\lambda$, $\alpha_{1}$ and $\alpha_{2}$. Consequently, we found theoretically that the zigzag model provides a fertile platform to model classically so-called displaced squeezed number states, being the light intensity distribution equivalent to the photon number distribution of such states. Finally, a freely available simulation tool to reproduce numerical results for any value of $\lambda$, $\alpha_{1}$ and $\alpha_{2}$ was published separately on nanoHUB platform \cite{32}.

\section{Acknowledgment}
B.M. Villegas-Martínez wish to express his gratitude to CONACyT, as well as to the National Institute of Astrophysics, Optics and Electronics INAOE for financial support.

\appendix 
\section{Exponential factorization of  $\exp\left( i \eta Z \hat{H} \right)$ }
We define auxiliary function $\hat{O}(Z)$ in terms to the exponential sum of operators $\hat{O}(Z)=\exp \left[i \eta  Z \left(\hat{K}^+ +\chi \hat{K}^0  +\hat{K}^-\right)\right]$, and its factorization $\hat{O}(Z)=\exp\left(i f \hat{K}^+ \right) \exp\left(i g \hat{K}^0 \right) \exp\left(i h \hat{K}^- \right)$. Differentiating  both equation with respect to $Z$ to give
\begin{equation} \label{ap01}
\frac{d\hat{O}(Z)}{dZ}=
i \left\lbrace  \left[ \frac{df}{dZ}   -i f \frac{dg}{dZ} - f^2 \frac{dh}{dZ} \exp\left(-i g \right)\right]  \hat{K}^+
+ \left[ \frac{dg}{dZ}-2 i f \frac{dh}{dZ} \exp\left(-i g \right) \right] \hat{K}^0
+  \frac{dh}{dZ} \exp\left(-i g \right)\hat{K}^- 
\right\rbrace \hat{O}(Z),
\end{equation}
\begin{equation}
\frac{d\hat{O}(Z)}{dZ}=i    \left(\eta\hat{K}^+ +\eta\chi \hat{K}^0  +\eta\hat{K}^-\right)\hat{O}(Z).
\end{equation}
Equating the coefficients of these two expressions yield to the differential equations system
\begin{subequations}\label{ap02}
\begin{align}
\frac{df}{dZ}+ \eta f^2 -i \eta\chi f &=\eta,
\\ 
\frac{dg}{dZ}-2 i \eta f &=\eta\chi,
\\
\frac{dh}{dZ} \exp\left(-i g \right)&=\eta,
\end{align}
\end{subequations} 
subjected to the initial conditions $f(0)=g(0)=h(0)=0$. We now distinguish two; the first one when  $\chi\neq \pm 2$, and the second one when $\chi=\pm 2$.

\subsection{Case $\chi\neq \pm 2$}
If $\chi\neq\pm 2$, the solution of above system of equations is given by  
\begin{subequations}\label{ap03}
	\begin{align}
f(Z)&=\frac{2 i}{\chi +i \sqrt{\chi ^2-4} \cot \left(\frac{1}{2} \eta Z \sqrt{\chi ^2-4} \right)},
\\
g(Z)&=2 \arctan \left[\frac{\chi  \tan \left(\frac{1}{2} \eta Z \sqrt{\chi ^2-4} \right)}{\sqrt{\chi ^2-4}}\right]
-i \left\lbrace \ln \left(4-\chi ^2\right)-\ln \left[2-\chi ^2+2 \cos \left(\eta Z \sqrt{\chi ^2-4} \right)\right]\right\rbrace,
\\
h(Z)&=f(Z).
\end{align}
\end{subequations}
$g(Z)$ can be simplified by assuming $\arctan(y)=-\frac{i}{2} \ln (\frac{1+iy}{1-iy})$ in the first term whereas in the second term we use $\ln\left(\frac{a}{b}\right)=\ln(a)-\ln(b)$ and $\cos(2 x)=2\cos^2(x)-1$ which leads to
\begin{equation} \label{ap04}
g(Z)=-i \Bigg\lbrace  \ln \left[\frac{\sqrt{\chi ^2-4} \cos \left(\frac{1}{2} \eta Z \sqrt{\chi ^2-4} \right) + i \chi  \sin \left(\frac{1}{2} \eta Z \sqrt{\chi ^2-4} \right) }{ \sqrt{\chi ^2-4} \cos \left(\frac{1}{2} \eta Z \sqrt{\chi ^2-4} \right) -i \chi  \sin \left(\frac{1}{2} \eta Z \sqrt{\chi ^2-4} \right) } \right]  +  \ln \left[\frac{\chi^2-4}{\chi^2-4 \cos^2 \left(\frac{1}{2} \eta Z \sqrt{\chi ^2-4} \right)}\right]   \Bigg\rbrace,
\end{equation}
by multiplying the denominator and numerator of the first term inside of logarithm with $\sqrt{\chi ^2-4} \cos \left(\frac{1}{2} \eta Z \sqrt{\chi ^2-4} \right) -i \chi  \sin \left(\frac{1}{2} \eta Z \sqrt{\chi ^2-4} \right)$ and after of it, we use the logarithmic relationship, $\ln\left(a b\right)=\ln(a)+\ln(b)$, to obtain
\begin{equation} \label{ap05}
g(Z)=-i \ln  \left[\cos \left(\frac{1}{2} \eta Z \sqrt{\chi ^2-4} \right) -i \frac{\chi}{\sqrt{\chi ^2-4}}   \sin \left(\frac{1}{2} \eta Z \sqrt{\chi ^2-4} \right)\right]^{-2}.
\end{equation}

\subsection{Case $\chi=\pm 2$}
If $\chi=\pm 2$, we must solve the equation $\frac{df}{dZ}+ \eta f^2 \mp 2 i \eta f =\eta$ to get
\begin{equation} \label{ap06}
f(Z)=\frac{\eta Z}{1 \mp i\eta Z} 
\end{equation}
With this value of the $f(Z)$ function, we solve the equation $\frac{dg}{dZ}=2 i \eta f \pm 2\eta$, which leads to
\begin{equation} \label{ap07}
g(Z)=\pm \pi+2i\ln\left(\pm i + \eta Z\right) 
\end{equation}
Using finally the equation $\frac{dh}{dZ} \exp\left(-i g \right)=\eta$, we obtain $h(Z)=f(Z)$.\\
Making the substitution of $\eta=2\alpha_2$ and $\chi=\frac{\lambda}{\alpha_2}$ in $f(Z)$ and $g(Z)$ of both cases and after some algebraic simplifications, it is possible to prove the following summary results
\begin{equation} \label{ap08}
f(Z)=
\begin{cases}
\frac{2 \alpha_{2}\sinh\left(\Gamma Z\right)}{\Gamma \cosh\left(\Gamma Z\right) - i \lambda  \sinh\left(\Gamma Z\right)},
& \lambda \neq \pm 2 \alpha_2 \\
\frac{2\alpha_2 Z}{1 \mp 2i\alpha_2 Z}, & \lambda = \pm 2 \alpha_2 \\
\end{cases}
\end{equation}
and
\begin{equation} \label{ap09}
g(Z)=
\begin{cases}
2 i \ln \left[\cosh \left(\Gamma Z\right) -i \frac{\lambda}{\Gamma} \sinh \left(\Gamma Z\right) \right],
& \lambda \neq \pm 2 \alpha_2 \\
2i\ln\left(1 \mp 2i\alpha_2 Z\right) , & \lambda = \pm 2 \alpha_2 \\
\end{cases}
\end{equation}
where
\begin{equation} \label{ap10}
\Gamma=\sqrt{4\alpha_2^2-\lambda^2}.
\end{equation}
Notice that the obtained expressions for the case $\chi=\pm 2$ are limit case when $\lambda \rightarrow \pm 2 \alpha_2$ in Eqs for $\chi\neq \pm2$.

\section{Matrix elements of the exponential operator $\exp\left( 2 i \alpha_{2}\hat{H} Z \right)$}
The matrix elements of $S_{m,k}$ can be calculate in a straightforward way by using the factorization form $\exp\left( 2 i \alpha_{2}\hat{H} Z \right)=\exp\left( g_1 \hat{K}^+ \right) \exp\left(g_0 \hat{K}^0 \right) \exp\left(g_1 \hat{K}^- \right) $ and introducing two identity operators between the exponential operators as
\begin{align}\label{B1}
S_{m,k}&=\bra{m} \exp\left( g_1 \hat{K}^+ \right) \sum_{j_1=0}^{\infty}\ket{j_1}\bra{j_1} \exp\left(g_0 \hat{K}^0 \right) \sum_{j_2=0}^{\infty}\ket{j_2}\bra{j_2} \exp\left(g_1 \hat{K}^- \right)  \ket{k}
\nonumber \\
&=\sum_{j_1,j_2=0}^{\infty} \bra{m} \exp\left( g_1 \hat{K}^+ \right) \ket{j_1}\bra{j_1} \exp\left(g_0 \hat{K}^0 \right) \ket{j_2}\bra{j_2} \exp\left(g_1 \hat{K}^- \right)  \ket{k},
\end{align}
which may be solved separately. For $\bra{j_2} \exp\left(g_1 \hat{K}^- \right)\ket{k}$ we have
\begin{align} \label{B2}
\bra{j_2} \exp\left(g_1 \hat{K}^- \right)  \ket{k}=
\begin{cases}
0, & j_2>k,\\
0, &  k-j_2 \text{ is odd,}\\
\frac{1}{\left(\frac{k-j_2}{2} \right)! } \sqrt{\frac{k!}{j_2!}}\left(\frac{g_1}{2} \right)^{\frac{k-j_2}{2}} ,  &  j_2\leq k \text{ and }  k-j_2 \text{ is even.}
\end{cases}
\end{align}
If we introduce the step function (which almost is the Heaviside function, but defined in 0 as 1)
\begin{equation} \label{B3}
\Theta(x)=
\begin{cases}
0, & x<0,\\
1, & x \geq 0
\end{cases}
\end{equation}
and we notice that  $\cos^2\left[\left(m-n \right)\frac{\pi}{2}  \right] $ is zero if $m$ and $n$ have different parity and 1 if they have the same parity, we can write
\begin{align} \label{B4}
\bra{j_2} \exp\left(g_1 \hat{K}^- \right)  \ket{k}=
\Theta(k-j_2)\cos^2\left[\left(k-j_2 \right)\frac{\pi}{2}  \right]\frac{1}{\left(\frac{k-j_2}{2} \right)! } \sqrt{\frac{k!}{j_2!}}\left(\frac{g_1}{2} \right)^{\frac{k-j_2}{2}}.
\end{align}
In the case for $\bra{j_1} \exp\left(g_0 \hat{K}^{0} \right)\ket{j_2}=\exp\left[\frac{g_0}{2} \left(j_2+\frac{1}{2}\right)\right] \delta_{j_1,j_2}$ and the solution for $\bra{m} \exp\left( g_1 \hat{K}^+ \right) \ket{j_1}$ is given by
\begin{align} \label{B6}
\bra{m} \exp\left( g_1 \hat{K}^+ \right) \ket{j_1}=
\begin{cases}
0, & m<j_1,\\
0, &  m-j_1 \text{ is odd,}\\
\frac{1}{\left(\frac{m-j_1}{2} \right)! } \sqrt{\frac{m!}{j_1!}}\left(\frac{g_1}{2} \right)^{\frac{m-j_1}{2}} ,  &  m\geq j_1 \text{ and }  m-j_1 \text{ is even.}
\end{cases}
\end{align}
which can be written as
\begin{align} \label{B7}
\bra{m} \exp\left( g_1 \hat{K}^+ \right) \ket{j_1}=
\Theta(m-j_1)\cos^2\left[\left(m-j_1 \right)\frac{\pi}{2}  \right]\frac{1}{\left(\frac{m-j_1}{2} \right)! } \sqrt{\frac{m!}{j_1!}}\left(\frac{g_1}{2} \right)^{\frac{m-j_1}{2}},
\end{align}
Substituting $\bra{j_1} \exp\left(g_0 \hat{K}^{0} \right)\ket{j_2}$ in the last line of \eqref{B1}, we obtain
\begin{align} \label{B8}
S_{m,k}&=\bra{m} \exp\left( 2 i \alpha_{2}\hat{H} Z \right) \ket{k}
=\sum_{j_1,j_2=0}^{\infty} \bra{m} \exp\left( g_1 \hat{K}^+ \right) \ket{j_1}
\exp\left[\frac{g_0}{2} \left(j_2+\frac{1}{2}\right)\right] \delta_{j_1,j_2}
\bra{j_2} \exp\left(g_1 \hat{K}^- \right)  \ket{k}
\nonumber \\  &
=\sum_{j=0}^{\infty} \bra{m} \exp\left( g_1 \hat{K}^+ \right) \ket{j}
\exp\left[\frac{g_0}{2} \left(j+\frac{1}{2}\right)\right] 
\bra{j} \exp\left(g_1 \hat{K}^- \right)  \ket{k},
\end{align}
and finally
\begin{align}  \label{B9}
S_{m,k}=&\bra{m} \exp\left( 2 i \alpha_{2}\hat{H} Z \right) \ket{k}
\nonumber \\
=&\sqrt{m!k!}\left(\frac{g_1}{2} \right) ^{\frac{m+k}{2}} \exp\left(\frac{g_0}{4} \right) 
\sum_{j=0}^{\infty} \frac{\Theta\left(m-j \right) \Theta\left(k-j \right) }{\left(\frac{m-j}{2} \right) ! \left(\frac{k-j}{2} \right) ! j!}  
\cos^2\left[\left(m-j \right)\frac{\pi}{2}  \right] \cos^2\left[\left(k-j \right)\frac{\pi}{2}  \right] 
\nonumber \left[ \frac{2}{g_1} \exp\left(\frac{g_0}{2} \right) \right]^j.
\end{align}

\end{document}